\documentclass[onecolumn]{aastex63}
\usepackage{xcolor}
\usepackage{hyperref}

\newcommand{\PSUAA}{Department of Astronomy \& Astrophysics, 525 Davey Laboratory, The Pennsylvania State University, University Park, PA, 16802, USA}
\newcommand{\PSUCEHW}{Center for Exoplanets and Habitable Worlds, 525 Davey Laboratory, The Pennsylvania State University, University Park, PA, 16802, USA}
\newcommand{\PSETI}{Penn State Extraterrestrial Intelligence Center, 525 Davey Laboratory, The Pennsylvania State University, University Park, PA 16802, USA}

\received{21 Jun 2021 }
\revised{10 Sep 2021}
\accepted{17 Sep 2021}
\submitjournal{the Astronomical Journal}

\usepackage{amsmath}
\usepackage{amsfonts}
\usepackage{amssymb}
\usepackage{upgreek}
\usepackage{graphics}
\usepackage{}

\usepackage{natbib}
\graphicspath{{./}{figures/}}

\begin{document}

\title{Stellar Gravitational Lens Engineering for Interstellar Communication and Artifact SETI}


\author[0000-0003-2633-2196]{Stephen Kerby}
\affiliation{\PSUAA}

\author[0000-0001-6160-5888]{Jason T. Wright}
\affiliation{\PSUAA}
\affiliation{\PSUCEHW}
\affiliation{\PSETI}

\begin{abstract}
    Several recent works have proposed ``stellar relay" transmission systems in which a spacecraft at the focus of a star's gravitational lens achieves dramatic boosts in the gain of an outgoing or incoming interstellar transmission. We examine some of the engineering requirements of a stellar relay system, evaluate the long-term sustainability of a gravitational relay, and describe the perturbations and drifts that must be actively countered to maintain a relay-star-target alignment. The major perturbations on a relay-Sun-target alignment are the inwards gravity of the Sun and the reflex motion of the Sun imparted by the planets. These $\sim\rm{m/s/yr}$ accelerations can be countered with modern propulsion systems over century-long timescales. This examination is also relevant for telescope designs aiming to use the Sun as a focusing element. We additionally examine prospects for an artifact SETI search to observe stellar relays placed around the Sun by an extraterrestrial intelligence and suggest certain nearby stars that are relatively unperturbed by planetary systems as favorable nodes for a stellar relay communications system.
\end{abstract}


\section{Introduction} \label{sec:intro}

Reliable and consistent direct communication over interstellar distances is a daunting challenge to modern engineering. An interstellar communication requires extraordinarily high directional gain or transmitting power to ensure that information packets can be reconstructed after steaming through deep space. Over the past forty years, various proposals have suggested using the gravitational lensing effect of a star like our Sun to focus outgoing or incoming transmissions, dramatically supplementing the gain of a transmitting relay spacecraft \citep{Eshleman, Maccone2011, Hippke2020i}.  A similar arrangement, like that proposed in the FOCAL mission, can focus photons from the distant universe onto a space telescope, dramatically increasing the magnification and effective collecting area of the system \citep{Focal2010}. Using the gravitational lens of a star, interstellar communication becomes feasible with only modestly-sized radio transmitters, indicating that it is theoretically possible for a small probe sent to a nearby star to use that star's gravitational lens to send transmissions back to its home planet.

A stellar relay obtains dramatically increased directional transmission gain using the gravitational lens of a star, and therefore emits the initial unfocused transmission towards the lensing star or a segment of its Einstein ring.  The collimated outgoing beam can have overall directional gains approaching 120 dB \cite{Maccone2011}, allowing for interstellar communication without inordinately high transmission power. If a stellar relay were using our Sun to focus its transmission to a target star, observers on Earth could intercept communications from the target star to the relay, or, by observing the point on the sky opposite the target star, detect transmissions from the relay spacecraft towards the inner solar system.

A historical and ongoing subfield of artifact SETI (Search for Extra-Terrestrial Intelligence) is the search for probes or spacecraft from an extra-terrestrial intelligence (ETI) in our solar system \citep{Bracewell1960}. An extraterrestrial spacecraft might remotely explore or monitor our solar system, but must maintain communications across interstellar distances to be scientifically productive.  Alternatively, given the daunting challenges of direct interstellar communication, a broadcasting spacecraft using the gravitational lens of the Sun as described in \cite{Maccone2011} and \cite{Hippke2020i} could facilitate interstellar communication as part of a broader communication network in which the Sun is just one node. A SETI search for ETI-placed stellar relay probes around the Sun was proposed in \cite{Gillon2014}.

Interestingly, the use of stellar relays by an ETI invokes would mean there could be copious ETI communications throughout the galaxy, but very few targets from which we could detect signals due to the very narrow outgoing beams.  A galactic civilization using stellar relays to tightly collimate radio beams for communication could result in none of those transmissions intersecting Earth, creating a seemingly silent cosmos \citep{Gertz2018}.

\begin{figure*}
    \centering
    \includegraphics[width=0.9\textwidth]{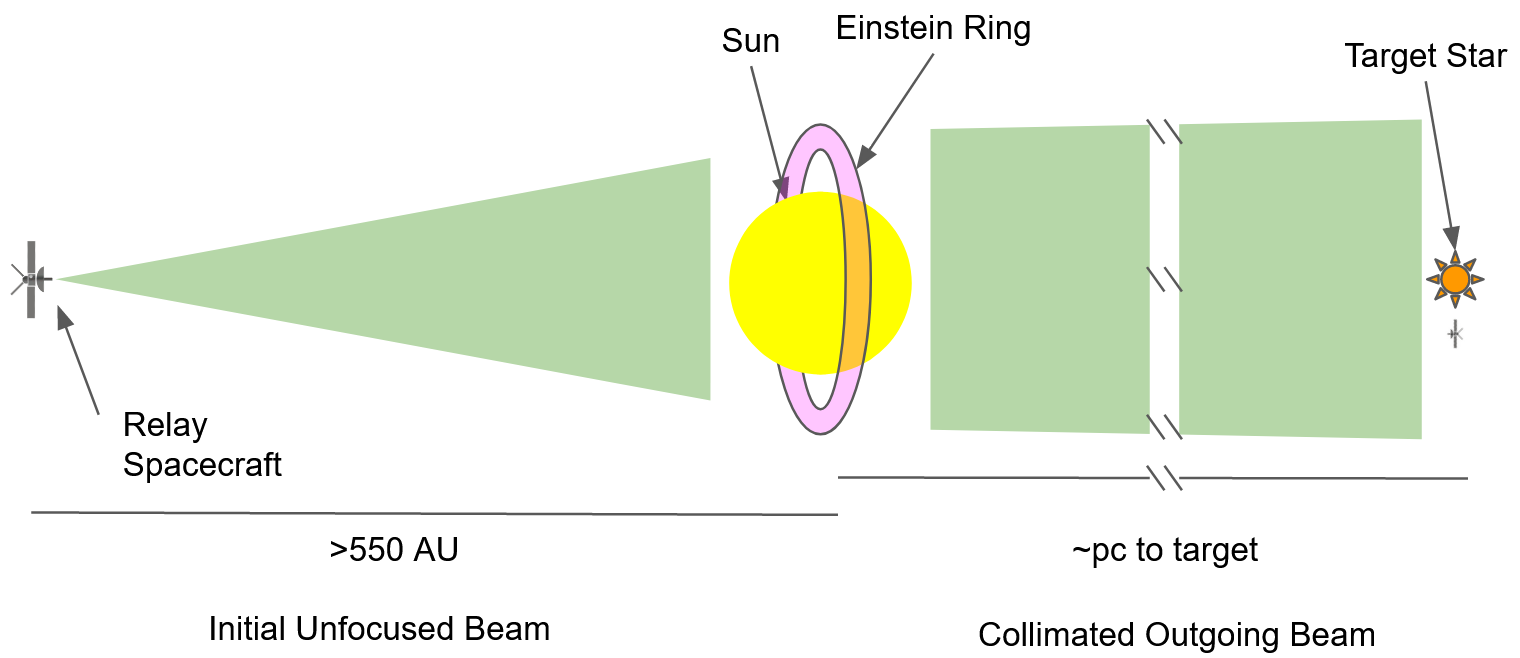}
    \caption{A schematic of a on-axis stellar relay transmission system, opening angles, distances, and sizes not to scale. The initial unfocused transmission beam may even have an annular pattern to prevent flux from being lost to the disk of the Sun. A reversed arrangement can be used to receive signals from a distant star by focusing rays onto the spacecraft. With a total system gain of $\sim 120$ dB, the outgoing beam can have an opening angle approaching $1 \arcsec$ as described in Section \ref{sec:Theory}, hundreds of times finer than the typical signal cone of a radio dish.}
    \label{fig:examplediagram}
\end{figure*}

A stellar relay can be designed to use the entire Einstein ring of the Sun if the relay spacecraft is on-axis, or transmit at only a single point on the Einstein ring. For simplicity we only consider on-axis broadcasting using the entire annular Einstein ring in this work, though an off-axis arrangement using only part of the Einstein ring follows similar proofs and obtains similar results \citep{Hippke2020i}. Figure \ref{fig:examplediagram} shows a schematic overview of a characteristic on-axis stellar relay using the Sun's entire Einstein ring.

Though recent work has established the theoretical feasibility of using stellar relays for communication \citep{Maccone2014,Hippke2020i}, there are outstanding engineering challenges that must be overcome to realize a long-lived stellar relay.  The difficulty of sustaining a long-lived stellar relay also informs the likelihood that an ETI might be currently using the Sun for a stellar relay, and therefore the probability of success of an artifact SETI search.

The main goal of this paper is to discuss several dynamical perturbations on a stellar relay, and examine possible methods that might be used to maintain a relay-lens-target alignment over extended periods of time. We compare the usefulness of the Sun as a stellar relay in a communications network against other nearby stars and examine how modern human spacecraft propulsion technologies might fare in sustaining a stellar relay. Finally, we postulate on whether artifact SETI searches for stellar relays around the Sun have reasonable hopes for success.

In section \ref{sec:Theory} we build a quantitative picture of a stellar relay from previous works and discuss details of a stellar relay transmission scheme. In section \ref{sec:Perturbations} we discuss dynamical perturbations on the relay-Sun-target syzygy (threefold alignment) and generalize those perturbations to any generic planetary system.  In section \ref{sec:artifactSETI} we examine artifact SETI considerations including the requirements on a stellar relay for its transmissions to be observable from Earth and the prospects of observing an extraterrestrial spacecraft acting as a stellar relay transmitter in the solar neighborhood. Finally, in section \ref{sec:Summary} we summarize our findings.

\section{Theory of the Stellar Gravitational Lens as a Focusing Element}
\label{sec:Theory}

A stellar relay has two components that must be managed and maintained: a host star and a relay spacecraft at distance $d$ from the host star. While this discussion is centered on using the gravitational lens of a star to focus outgoing transmissions, similar calculations apply to a telescope using the star's lens to collect light \citep{Turyshev2018}. For a transmitting spacecraft, the directional gain\footnote{Gain is interchangeably given in absolute units or in decibels, with $G(\rm{dB}) = 10 \log G$} of a transmission dish of area $A_{\mathrm{SC}} = \pi R_{\mathrm{SC}}^2$ and efficiency $\epsilon_{\mathrm{SC}}$ transmitting at wavelength $\lambda$ is given by

\begin{equation}
    G_{\rm{SC}} =  \frac{4\pi^2 R_{\rm{SC}}^2 \epsilon_{\rm{SC}}}{\lambda^2} \approx 3.9 \times 10^5 \epsilon_{\rm{SC}} \left( \frac{R_{\rm{SC}}}{\rm{m}} \right)^2 \left( \frac{\lambda}{\rm{cm}} \right)^{-2}
\end{equation}

\noindent for monochromatic transmission. Once the initial unfocused beam is produced by the relay spacecraft, the transmission streams towards the host star. 

Gravitational lensing is the process by which the path of light is bent around massive objects like black holes, stars, or planets. General relativistic calculations show that the minimum focal distance, the distance at which the Einstein ring of an distant circular object is just outside the object's disk, is dependent on the object's physical radius $R_{\star}$ and Schwarzschild radius $R_s$ with

\begin{equation}
    d_{\rm{focal}} > \frac{R_{\star}^2}{2R_s}
\end{equation}

\noindent which for the Sun is $550\:\rm{AU}$ \citep{Maccone2011}. Geometrically speaking, all points beyond this distance serve equally well as the focus for a target on the exact opposite side of the sky with respect to the Sun. However, several competing factors restrict a relay spacecraft to a certain band of distances from the gravitational lens. If the transmitter is too distant from the Sun, the initial transmission suffers from distance losses. If a transmission is aimed too close to the Sun's disk, the solar corona interferes with radio transmissions while contributing a bright background of photons. A relativistic treatment \citep{Orta1994} of the gravitational lens of the Sun gives a solar gain that is linearly related to the Schwarzschild radius of the Sun

\begin{equation}
    G_{\odot} = \frac{4\pi^2 R_S}{\lambda} \approx 1.2 \times 10^7 \left( \frac{M_{\rm{lens}}}{M_\odot} \right) \left( \frac{\lambda}{\rm{cm}}\right)^{-1}
\end{equation}

\noindent Clearly, including the Sun's gravitational focusing effect in an interstellar communications link can increase the gain of the system by around $70\:\rm{dB}$ if the entire Einstein ring is targeted by the initial transmission. The total gain of the stellar relay system is simply the product of the spacecraft and solar gains, plus another efficiency factor $\epsilon_e$ to account for the transmission flux that is not focused into the final outgoing beam. This total system gain is given by

\begin{equation}
    G_{\rm{tot}} = \epsilon_e G_\odot G_{\rm{SC}} = \epsilon_e \frac{4\pi^2 R_S}{\lambda} \frac{4\pi^2 R_{\rm{SC}}^2 \epsilon_{\rm{SC}}}{\lambda^2}
\end{equation}

\begin{equation}
    G_{\rm{tot}} \approx 4.7 \times 10^{12} \epsilon_{\rm{SC}} \epsilon_e \left( \frac{M_{\rm{lens}}}{M_\odot} \right) \left( \frac{\lambda}{\rm{cm}}\right)^{-3} \left( \frac{R_{\rm{SC}}}{\rm{m}} \right)^2
\end{equation}

\noindent which shows that a stellar relay using the gravitational lens of the Sun and a modest meter-scale radio transmission dish could produce gains above $120\:\rm{dB}$ for $\lambda<\rm{cm}$. For this work, we write equations using wavelength in terms of $\rm{cm}$, despite the deflection of radio waves by the solar corona at these wavelengths \citep{Hippke2020i}.  Centimeter and longer wavelengths suffer significant degradation of focus due to the corona, signficantly reducing the gain of the lens. In this work, our use of centimeter for wavelengths reflects the longest wavelengths at which the lens will be useful.

\subsection{Off-Axis Gain Pattern}

To motivate the need for precise positioning of the relay spacecraft, it is worthwhile to examine the difficulties of receiving signals focused by the solar gravitational lens (SGL) instead of transmitting via the same. Why must a spacecraft using the gravitational lens of a star maintain precise positioning at all?  The need for precise stationkeeping is informed by the incoming or outgoing flux distribution produced by the gravitational lens.

The total gain described above assumes that the relay-Sun-target alignment is absolutely exact, and the Sun is functioning as a perfectly spherical mass distribution. Assuming the Sun functions as a spherical mass, \cite{Turyshev2003} describes the off-axis flux pattern at the image plane of the SGL as a 0th-order Bessel function. The total gain of the system is modulated by this Bessel function, a function of distance to the alignment axis $b$. However, incorporating the Sun's oblate spheroidal shape and quadrupole moment, \cite{Loutsenk2017} derives the flux pattern of the Sun as lensing element.  Their proof shows that instead of focusing incoming light onto a ideal point focus, an oblate spheroid like the Sun produces a four-pointed flux pattern beyond the minimum focal distance. Equation 17 in \cite{Loutsenk2017} gives the diameter of this flux pattern, which informs the positional stability needed for maintaining alignment with an interstellar target. Using our notation conventions,

\begin{equation}
    \Delta X = 4 I_2 R_\odot \sqrt{\frac{550 \:\rm{AU}}{d}} \sin^2 \beta
\end{equation}

\noindent with $I_2 \approx 2 \times 10^{-7}$ is the Sun's quadrupole moment and $\beta$ the viewing angle with respect to the Sun's polar axis.  Using $\beta = \pi/2$ the size of the flux pattern at $550 \:\rm{AU}$ is $\sim 500 \:\rm{m}$.  The massive gains that would normally be granted by an ideal point focus are instead spread across a region of approximate size $\Delta X$, and the high gain estimates described by \cite{Maccone2014} are an overestimate. To maintain a stable lock on the target on the opposite side of the Sun, the positional accuracy of a relay spacecraft is therefore of order $\sim 100\:\rm{m}$.

Uncontrolled drift through the edges of the flux pattern would result in rapidly fluctuating transmission strengths, suggesting that a receiver spacecraft must actively and autonomously correct its location to maintain alignment while transmitting. Clearly, a primary engineering challenge in building a stellar relay system is autonomously maintaining sub-$\rm{km}$ positioning precision over practically interstellar distances. Using the Einstein ring image of the target star or background objects via an optical monitoring telescope can serve as a useful guide to maintaining this high degree of precision.

\subsection{Outgoing Beam Opening Angle}

An outgoing interstellar transmission using the SGL must be aimed precisely at the target star across interstellar distances, and the incredibly high gains obtained using the lens of the Sun requires precise aiming given the small beam opening angle. The solid angle covered by a beam emerging from a transmission of gain $G$ is defined as

\begin{equation}
    \Omega_A = \frac{5.3 \times 10^{11} \square \arcsec}{G}
\end{equation}

\noindent with $\Omega_A$ in square arcseconds ($\square \arcsec$). As the total gain approaches $10^{12}$, the area on the sky covered by the beam drops to less than a square arcsecond. 

For a stellar relay, $G_{\rm{tot}}=G_{\rm{SC}}G_\odot$. The target star must constantly be kept in the patch of sky covered by the outgoing focused beam by active stationkeeping by the relay spacecraft to maintain the transmission link. Solving for the opening angle in arcseconds $\pi \theta_A^2 = \Omega_A$ gives an expression for the degree of precision that must be maintained in the relay-Sun-target alignment to ensure the transmission is received.

\begin{equation} \label{eq:thetaA}
    \theta_A = \sqrt{\frac{5.3 \times 10^{11} \square \arcsec}{\pi G_{\rm{tot}}}}
\end{equation}

The alignment between the relay probe, the Sun, and the target star must be maintained to within this tolerance to maintain a communication link. For a probe at distance $d$ from the Sun, the physical distance tolerance is $\delta_A \approx d \theta_A$. A gain around $120 \:\rm{dB}$ with $d \approx 550 \:\rm{AU}$ translates to a physical tolerance in spacecraft position of $\delta_A \approx 0.001 \:\rm{AU} \approx 150000 \:\rm{km}$. If the spacecraft or the Sun moves off-axis by an angular distance large enough to break this alignment, the spacecraft must fire its thrusters to compensate and prevent a loss of signal. Fortunately, a target star will appear as a bright Einstein ring around the Sun as its light is focused by the gravitational lens, allowing for active correction of deviations from perfect trifold syzygy.

\section{Perturbations on Relay-Sun-Target Alignment}
\label{sec:Perturbations}

In this section, we discuss the perturbations that might disturb a relay-Sun-target alignment.  These perturbations can be extended to other planetary systems with ease, but in the foreseeable future the only planetary system of interest for stellar relay technology is our own.  The most significant of these perturbations are the inwards gravitational force on the relay by the target star and the reflex motion of the target star by orbiting celestial objects. For a relay designed for continual transmission to the target star, dynamical perturbations require propulsive correction by the spacecraft to keep the target star along the relay-Sun vector and beyond the minimum focal distance. 

Secondary perturbations include the proper motion of the target star viewed from interstellar distances and the orbital motion of the receiver viewed across interstellar distances. The proper motion of the target star is a small linear effect, with most stellar proper motions not exceeding a few arcseconds per year. To first-order, the proper motion of the target must be countered by a purposeful drift in the relay alignment. To second-order, perspective changes create slight proper accelerations, but these are less than $10^{-10} \:\rm{m/s/s}$ for the nearest stars, far below other terms in our analysis. Differential galactic acceleration between the Sun and nearby stars is many orders of magnitude smaller. 

While the receiver spacecraft may also be moving in an orbit of a distant star or planet \citep[as discussed in][]{Hippke2021}, its motion is viewed across an interstellar distance $D$. For example, Proxima Centauri ($D = 270000 \:\rm{AU}$) has a planet at $a = 1.5 \:\rm{AU}$ \citep{Damasso2020}, which would have a relatively large angular orbital radius as viewed from Earth. Even in this extreme case, that planet's orbital motion only results in an angular displacement of about an arcsecond over its five-year orbit, a perturbation that can be neglected for all but the most focused transmission systems. If a spacecraft using the Sun's gravitational lens must maintain exact alignment with a receiver on a distant exoplanet, the corresponding acceleration of the transmitting spacecraft is reduced by a factor of $D/d$ compared to the orbital acceleration of the target exoplanet due to the central position of the Sun in the transmitter-Sun-target alignment.

Because of the tiny angular radius of exoplanet orbits viewed from Earth (rarely exceeding $1 \arcsec$) with the small cost of countering the orbital acceleration of a distant receiver, the orbital motion of a receiver spacecraft across interstellar distances is much smaller than the other perturbations considered here.

Even a relay designed only for intermittent transmission will need to adjust its position to a similar degree to ensure a stable transmission link at the time of transmission. It is infeasible for the relay to opportunistically lurk unmoving in interstellar space and only transmit when the Sun swings into position, as the transmission would have unpredictable total gain and would only occur at chance intervals. Additionally, a transmitting spacecraft must ``lead" its target's proper motion through space to account for the travel time of the transmission across the intervening distance $D$.

\subsection{Inwards Gravitational Force}

A relay positioned at $d\sim 600 \:\rm{AU}$ from the Sun sits between the Kuiper belt at $40 \:\rm{AU}$ and the Oort cloud at at $> 2000 \:\rm{AU}$. At $d > 550 \:\rm{AU}$, the gravitational field of the Sun is quite small, but not insignificant over long timescales.  The gravitational acceleration of the Sun at some distance $d$ is

\begin{equation}
    A_G = \frac{GM_\odot}{d^2} \approx 2 \times 10^{-8} \left( \frac{d}{550 \:\rm{AU}} \right)^{-2} \:\rm{m/s/s} \approx 0.62 \left( \frac{d}{550 \:\rm{AU}} \right)^{-2} \:\rm{m/s/yr}
\end{equation}

While the magnitude of this acceleration is very small, a relay designed to remain stationary at the solar focus for longer than the free-fall time must pay this delta-v\footnote{delta-v, normally measured in $\rm{m/s}$, describes the velocity change available to a spacecraft using propulsive fuel.} for its expected mission lifetime. If this stationkeeping debt is not paid, the relay will fall into the Sun over $\sim 5000$ years, entering a comet-like orbit.  This consideration suggests that a relay spacecraft that ceases to resist the Sun's gravity will be displaced, ejected, or destroyed as its orbit decays over a few millennia.

All other forces being absent, this gravitational force places an upper limit on the longevity of a stellar relay using active propulsion to stationkeep.  A relay spacecraft around the Sun can only last for lifetime in years $\tau_\mathcal{L}$ approximately equal to twice its available delta-v in $\rm{m/s}$ ($\Delta v$).  

\begin{equation}
    \tau_\mathcal{L} < \frac{\Delta v}{A_G} \approx 2 \:\rm{yr} \left( \frac{\Delta v}{\rm{m/s}} \right) \left( \frac{d}{550 \:\rm{AU}} \right)^{2}
\end{equation}

Naturally, this stationkeeping only applies an active cost to spacecraft using fuel for propulsion; NASA's interplanetary probes normally have a delta-v budget $\sim 1000 \:\rm{m/s}$, and so could pay this stationkeeping cost for thousands of years, much longer than the mechanical lifetime of modern probes or the free-fall time for a derelict spacecraft falling inwards from $550 \:\rm{AU}$. 

Finally, a spacecraft could persist without explicitly countering the gravity of the Sun by coasting along the line of the Sun's focus. A probe with close to zero gravitational binding energy with respect to the Sun could persist along this line for a very long time, coasting outwards or inwards until it suffers from distance losses or falls closer than the minimum focal distance. In the latter case, it might use a small amount of propellant to ensure it ``misses" the Sun and returns to the focal line on a long comet-like orbit. For stars with no orbital companions, such a scheme might require less propellant than minimum values discussed herein.

\subsection{Stellar Reflex Motion}

Any star with companion planets or in a binary system will have a reflex motion imparted by the orbital motion of its companions. This manifests as an acceleration on the star by the companion, generating reflex motion. A relay spacecraft hoping to stay exactly on-axis in the relay-star-target syzygy must mirror the motion and acceleration of the lensing star if such motion would break the target alignment or change the direction of the outgoing beam.

Notably, the reflex motion of the host star is only relevant if the outgoing focused beam is narrow enough such that the motion of the host star would swing the transmission beam off the target at interstellar distance. Assuming the companions mass has a circular orbit and ignoring inclination effects (a factor of $\sim 2$ at most), conservation of angular momentum gives the angular reflex displacement $\Phi_{\rm{ref}}$ for a star of mass $M_1$ with a companion body (mass $M_2$) separated by semi-major axis $a$ viewed from the relay spacecraft at distance $d$ as

\begin{equation}
    \Phi_{\rm{ref}} = \sin^{-1} \left( \frac{a}{d(1+\frac{M_1}{M_2})} \right)
\end{equation}

If the angular reflex motion amplitude of the host star $\Phi_{\rm{ref}}$ is greater than the opening angle of the outgoing beam, the spacecraft must actively correct its position to account for the displacement of the host star by the companion body. This limit is reached if the displacement of the host star by the reflex motion is equal to or greater than the opening angle of the outgoing beam given by equation \ref{eq:thetaA}.

The path the relay spacecraft must trace to maintain a perfect relay-Sun-target alignment is identical to the projected motion of the Sun due to the reflex motion. That path has the same period and magnitude as the motion of the Sun, and therefore the acceleration that the relay spacecraft must apply is conveniently the same as the gravitational force the companion imparts on the Sun. For a circular orbit, this acceleration is dependent on the companion mass $M_2$ (or its ratio with Jupiter's mass $M_J$) and the distance between the host star and the companion $a$.

\begin{equation}
    A_{\rm{ref}} = \frac{GM_2}{a^2} \approx 5.6 \times 10^{-6} \:\rm{m/s/s} \left( \frac{M_2}{M_J} \right) \left( \frac{a}{\rm{AU}} \right)^{-2} \approx 177 \:\rm{m/s/yr} \left( \frac{M_2}{M_J} \right) \left( \frac{a}{\rm{AU}} \right)^{-2}
\end{equation}

Table \ref{tab:Companions} compares the accelerations imparted on the Sun by real and hypothetical companions, the amplitude of the Sun's reflex motion due to those companions as viewed from $550 \:\rm{AU}$, and the lowest system gain at which each reflex motion can create a misalignment by drawing the Sun's gravitational lens off-axis from the target star.

\begin{deluxetable*}{cccccc}
\tablecaption{Real and hypothetical (marked with $\star$) reflex motions imparted on a Sun-like star. $\Phi_{\rm{ref}}$ is calculated assuming the relay spacecraft is positioned at $550\:\rm{AU}$. If a stellar relay has a system gain $G_{\rm{tot}} > G_{\rm{lim}}$, it must match the acceleration imparted on the Sun by its companion to maintain alignment by paying a delta-v cost $A_{\rm{ref}}$.}
\label{tab:Companions}
\tablewidth{\textwidth}
\tablehead{ 
\colhead{Companion} & \colhead{$M_2$} & \colhead{$a$} & \colhead{$\Phi_{\rm{ref}}$} & \colhead{$G_{\rm{lim}}$} & \colhead{$A_{\rm{ref}}$} \\
\colhead{} & \colhead{($M_J$)} & \colhead{($\rm{AU}$)} & \colhead{($\arcsec$)} & \colhead{(dB)} & \colhead{$(\rm{m/s/yr})$}
}
\startdata
None (solar gravity only) & & & & 0 & 0.62 \\
Earth & 0.003 & 1 & $1.1 \times 10^{-3}$ & 171 & 0.53 \\
Jupiter & 1 & 5.2 & $1.9$ & 106 & 6.6 \\
Saturn & 0.3 & 9.6 & $1.07$ & 112 & 0.58 \\
Neptune & 0.05 & 30.1 & $0.56$ & 117 & 0.01 \\
Hot Jupiter$^\star$ & 5 & 0.1 & $0.18$ & 128 & 89,000 \\
Brown Dwarf ($\sim \: $ GJ 229B)$^\star$ & 50 & 20 & $340$ & 62 & 22 \\
Close Large Star ($\sim \alpha $ Cen B)$^\star$ & 900 & 20 & $3500$ & 41 & 400 \\
Distant Small Star ($\sim$ Proxima Cen ) $^\star$ & 120 & 9000 & $3.6 \times 10^5$ & 1.4 & $2 \times 10^{-4}$ \\
\enddata
\end{deluxetable*}

\subsection{Comparative Stellar Relay Hospitality}

In the previous section, we demonstrated that Jupiter alone can require a relay spacecraft around the Sun to expend a few $\rm{m/s}$ of delta-v per year to stay aligned, while a hot Jupiter or other close-in massive companion can require extremely costly alignment maintenance. It is not unreasonable to expect that a Sun-like star without a Jupiter-mass planet orbiting would be a more economical host for a long-lived stellar relay, as the relay spacecraft would expend delta-v at a much lower rate. In a similar vein, a civilization hoping to build a galaxy-spanning stellar relay network can discriminate possible host stars based on the dynamical cost of maintaining the relay. This feature has implications for artifact SETI searches for stellar relay spacecraft, discussed below.

Not all perturbing planets make a stellar relay around a star impossible. For example, a ``hot Jupiter" does not displace the star by a large angle with respect to a relay spacecraft at $550 \:\rm{AU}$, so relays of moderate ($\sim 120 \:\rm{dB}$) total gain need not pay the hefty cost to maintain perfect alignment in that case; the maximum displacement of the star will not cause the outgoing beam to miss the target.  Likewise, a distant stellar companion like Proxima Centauri imparts a large angular displacement on the primary, but only over an extremely long time period, so the delta-v cost per year is small.

There are some factors that can disqualify a star from being an economical host for a constantly transmitting stellar relay network. For example, a relay spacecraft using $\alpha$ Cen B or Sirius A for gravitational focusing as suggested in \cite{Maccone2011} would orbit both stars in those binaries while only using the lensing effect of one. Due to the reflex motion imparted by $\alpha$ Cen A ($m_p = 1100 \: M_J$ and $r_p \approx 25 \:\rm{AU}$) or Sirius B ($m_p = 1000 \: M_J$ and $r_p \approx 20 \:\rm{AU}$) the spacecraft would need to expend hundreds of $\rm{m/s}$ each year to stay aligned, depleting a spacecraft propulsion system and causing the spacecraft to eventually drift out of position. A long-lived stellar relay around stars like $\alpha$ Cen B or Sirius B is significantly more costly to maintain than many other nearby stars. Additionally, the presence of a stellar-mass companion in contact binary would seriously complicate the geometry of the gravitational lens as well.

Clearly, stars with close stellar companions are unsuitable as stellar relays unless the system gain so low that the reflex motion of the star does not cause outgoing beam misalignment. Alternatively, the propulsion system on the relay spacecraft can be made both efficient and powerful to `brute force' the alignment problem. These problems could be overcome by paying the high delta-v cost to maintain alignment, or those binary star systems could simply be passed over during construction of a stellar relay network.

Additionally, the oblate spheroid shape of rotating stars deforms the gravitational lens away from an ideal circular lens. While no extensive treatment of the transmission flux pattern or gain possibilities has reached the level of detail of the receiver treatment in \cite{Loutsenk2017}, it is likely that rapidly rotating stars, centrifugally shaped, will serve as less ideal lenses for a stellar relay transmission system due to the smeared-out foci of such a lens.

With these considerations in mind, it seems that if interstellar engineering for a stellar relay must be economical in its choice of stellar relay host, there are dynamical guidelines that inform whether a star is suitable for long-duration stellar relay operation.

\begin{itemize}
    \item Close or moderately-distanced companion stars impart huge delta-v costs on a relay spacecraft, and thus close binary or multiple-star systems should be avoided.
    \item The presence of gas giant planets either limits the maximum gain total of the stellar relay (depending on the reflex semi-major axis imparted on the host star) or imparts a delta-v cost of a few $\rm{m/s}$ per year to maintain alignment.
    \item A more massive host star requires proportionally higher delta-v costs to resist its inwards gravity.
    \item A host star that is rapidly rotating will be deformed away from a spherical shape, resulting in a lens that does not have an exact focal point from which to transmit, resulting in significantly lowered gains.
\end{itemize}

By these measures, while the Sun is not in a close binary star system, its planetary system does impart reflex motions that warrant constant fine positional adjustments. The Sun also does not rotate with as much rapidity as many other stars, though it is significantly oblate nonetheless. A star with a less massive planetary system or lacking companions entirely would be more suited for a long-lived stellar relay than the Sun, especially if it is closer to spherically shaped. Only a complete stellar dynamics and exoplanet survey of nearby stars will reveal whether any ideal systems exist in the solar neighborhood.

\section{Artifact SETI Searches for Stellar Relay Spacecraft}
\label{sec:artifactSETI}

\subsection{Artifact SETI Prospects}

While there have been several proposals to use the SGL for telescope observations \citep{Focal2010,Maccone2011,Hippke2020i}, it is possible that a stellar relay could be leveraged by an extraterrestrial civilization to transmit reliably across interstellar distances. If an ETI has placed a stellar relay to use the Sun's gravitational lens for scientific exploration, transmissions to or from that spacecraft might be visible from Earth. Similarly, if an ETI building a galaxy-spanning communication network of relays uses dynamical factors to choose the most hospitable host stars, our discussion of the stability of a relay-Sun-target alignment might inform whether we expect such a relay to be present around the Sun. These economical considerations add to the nascent approach of using optical properties of gravitational lens transmission schemes to narrow down the spatial region suitable for hosting a transmitter \citep[for example ][]{Hippke2021}.

Naturally, interstellar transmission from a relay spacecraft would only be observable from Earth if the Earth is included in the initial unfocused beam. Even if the initial transmission is focused tightly on the Sun and thus invisible to Earth at most times, a relay spacecraft might broadcast other signals to other probes in the inner solar system. These secondary transmissions would be more likely to include Earth in their footprint, allowing the relay spacecraft to be observed nonetheless. Without these additional transmissions, the signal from the relay spacecraft could only be detected when the Earth's orbit takes it into a fortuitous four-way alignment with the Sun's Einstein ring, the spacecraft, and the target star, which can only occur for target stars along the ecliptic.

For interstellar transmissions from a relay spacecraft to be detectable from the Earth, the stellar relay must be actively transmitting.  A stellar relay that is constantly transmitting is more likely to be defected from Earth compared to a relay that only intermittently sends signals to the target star or local probes. An active relay spacecraft must have a sufficiently long-lived source of power to support almost continual transmission, and a propulsion system that can stationkeep at the focal point of the Sun. These systems must have mechanical integrity and fuel enough to support the spacecraft for millennia, the usual requirements for interstellar travel.

In a galactic transmission system that includes the Sun as a stellar relay host, there may also be transmissions approaching the solar system from other nearby stars.  These transmissions may be visible for terrestrial observers, and it may be productive to conduct radio SETI searches of nearby stars that lack perturbations from high-mass planets like Proxima, Barnard's Star, or Ross 154. An artifact SETI search for a stellar relay placed by an ETI might proceed by looking for radio transmissions from a point on the sky opposite these significant stars.  If a search aims to detect the primary interstellar transmission from the relay spacecraft, it would be more productive to limit searches to those times when the Earth joins a four-way relay-Earth-Sun-target alignment to ensure that the Earth is within the transmission cone of the initial unfocused beam. Such searches would be limited to stars at low ecliptic latitude.  

To qualitatively estimate the likelihood of an alien stellar relay being active around the Sun, it is useful to examine the different uses of a stellar relay.  Human scientists have already proposed leveraging the gravitational lens of the Sun for observational missions with space telescopes and for interstellar communications via a one-way or two-way relay bridge \citep{Maccone2008, Maccone2014, Eshleman1979, Turyshev2019}. An ETI could use stellar relay techniques for a similarly diverse range of goals, each goal informing the design considerations that the relay spacecraft would aim towards and the expected longevity of a particular stellar relay system. 

If an ETI is constructing a stellar relay for purely scientific exploration, the host star of the relay spacecraft would be uniquely selected among the stars in the galaxy for its scientific merit. Relay spacecraft placed for transmitting scientific findings need only remain active as long as exploration of the solar system continues. From an artifact SETI perspective, short-lived artifacts require extraordinary coincidence to still be active around the Sun, or else continuing scientific justification.

An ETI creating a communications network throughout the galaxy has a more vested economic interest in the longevity of individual relay spacecraft, and might design propulsion systems that can maintain relay spacecraft for millions of years. A node-based communications system might use stellar relays at only the most optimal hosts in the galaxy to launch transmissions over vast $\rm{kpc}$ spans, or most stars in the galaxy might have relays to form a vast galactic network. In that case, the likelihood of a relay being active around the Sun then depends on the extent of the galactic communications network. A node-based system with only a smattering of stellar relays in any galactic region is less likely to include a relay spacecraft around the Sun if the Sun is of middling friendliness due to dynamical reasons. 

\subsection{Human Spacecraft Stellar Relay Capabilities}

Artifact SETI searches normally focus on remnants of ETI activity that endure over timescales of millions of years or more. If the delta-v cost of aligning a long-lived stellar relay transmitter around the Sun is somewhere between $1$ and $10 \:\rm{m/s/yr}$, what sort of propulsion technology would ensure that a stellar relay could endure long after it was established, allowing success in artifact SETI searches? If we demand the relay spacecraft remain stable for millions of years, we likewise demand a propulsion system supplying at least several million $\rm{m/s}$ of delta-v.

Once the relay spacecraft exhausts its propellant, it will become derelict and drift out of position, reducing the likelihood that an artifact SETI search would be able to detect the spacecraft. Humanity's propulsive technology has expanded dramatically in a century of rocketry and spaceflight, encompassing chemical, nuclear fission, and electronic rockets and currently developing solar sail technology.  Many visions of interstellar travel and communication further posit the harnessing of nuclear fusion or antimatter propulsion \citep{Bracewell1960,Purcell1960}, approaches that could equally serve as propulsive and power generating systems onboard a spacecraft.  This range of technologies covers a whole range of efficiencies for a space probe, from propellant-less solar sails to primitive chemical rockets and fantastic nuclear engines.

The Tsiolkovsky rocket equation gives the amount of delta-v expected from a rocket system of exhaust velocity $v_{\rm{ex}}$, fully fueled mass $m_0$, and dry mass $m_e$ as

\begin{equation}
    \Delta v = v_{\rm{ex}} \ln{ \left( \frac{m_0}{m_e} \right)}
    \label{RocketEquation}
\end{equation}

\noindent which provides an estimate for the total propulsive capabilities of a spacecraft. The total lifetime of the probe is then

\begin{equation}
\tau = \frac{\Delta v}{A_\mathrm{ref}} = \frac{v_\mathrm{ex}}{A_\mathrm{ref}}\ln{ \left(1+\frac{m_\mathrm{fuel}}{m_e} \right)}
\label{Lifetime}
\end{equation}

Table \ref{tab:Spaceships} describes the exhaust velocity and delta-v expected from generic hypothetical vessels assuming that half the total mass of the relay spacecraft is devoted to fuel stores, $m_0/m_e = 2$ or $m_\mathrm{fuel}=m_e$. Modern orbital spacecraft are almost entirely propelled by chemical rockets and electric ion engines, though after the Apollo programs NASA built a prototype nuclear thermal rocket \citep[called NERVA, see][]{Ludewig2006} which used a nuclear reactor to superheat hydrogen as exhaust.  Futuristic modes of propulsion, such as nuclear fusion engines \citep{Genta2020} or antimatter reactors \citep{Homlid2020}, are included only with order-of-magnitude guesses for efficiency .

\begin{deluxetable*}{cccccc}
\tablecaption{Order-of-magnitude hypothetical spacecraft capabilities are calculated assuming half the spacecraft mass is devoted to fuel. Maximum lifetimes for a Solar relay spacecraft at $550 \:\rm{AU}$ are calculated considering only the gravity of the Sun ($\tau_{\rm{max}}$, using a cost of $0.5\:\rm{m/s/yr}$ in a situation where the spacecraft has low enough gain to ignore the perturbations by the planets) and in a high-gain system where the spacecraft must account for the perturbations of all the planets ($\tau_{\rm{all}}$, using  a cost of $ \sim 8\:\rm{m/s/yr}$). For these calculations we used the non-relativistic rocket equation, and therefore the estimates for high-energy propulsion systems are only order-of-magnitude.}
\label{tab:Spaceships}
\tablewidth{\textwidth}
\tablehead{ 
\colhead{Spaceship} & \colhead{Propellant} & \colhead{$v_{\rm{ex}}$} &  \colhead{$\Delta_V$} & \colhead{$\tau_{\rm{max}}$} &  \colhead{$\tau_{\rm{all}}$}\\
\colhead{} & \colhead{} & \colhead{($\rm{m/s}$)}  & \colhead{($\rm{m/s}$)} & \colhead{($\rm{yr}$)}  & \colhead{($\rm{yr}$)}
}
\startdata
Chemical Rocket & Fuel/Oxidizer & $3000$ &  $2080$ & $4160$ & $260$ \\
Nuclear Thermal Rocket & Hydrogen & $10000$  & $6930$ & $13860$ & $866$ \\
Xenon Ion Engine & Xenon & $50000$  & $34660$ & $69320$ & $4333$ \\
Continuous Nuclear Fusion Rocket & Deuterium & $4.5 \times 10^5$ &  $3.1 \times 10^5$ & $6.2 \times 10^5$ & $3.9 \times 10^4$ \\
Antimatter Rocket & Matter/Antimatter & $3.0 \times 10^6$ &  $2.1 \times 10^6$ & $4.2 \times 10^6$ & $2.6 \times 10^6$ \\
\enddata
\end{deluxetable*}

An ETI hoping to maintain a single stellar relay around the Sun for millions of years while paying a small but non-zero stationkeeping delta-v cost ($\sim$few m/s/yr for the Sun) must use advanced, highly efficient propulsion technology. Even modern ion engines do not have high enough exhaust velocity to provide $\sim 10^6 \:\rm{m/s}$ of delta-v. A relay spacecraft like NASA's Dawn mission, propelled by ion engines, would either need refueling or replacement after a few millennia, even if the exhaust velocity was increased by a factor of $10$. Notably, the LISA gravitational wave mission faces similar long-duration high-precision stationkeeping challenges as a stellar relay, using microthrusters that can provide $\sim 10^{-8} \:\rm{m/s/s}$ acceleration with autonomous control \citep{Jennrich2009}. Even those efficient electric propulsion systems could only resist the Sun's gravity and/or mirror the Sun's reflex motion for thousands, not millions, of years. Several modern and hypothetical methods of supplying propulsive forces for a relay spacecraft are described in Table \ref{tab:Spaceships}, including the operational lifetime of each given the perturbation on the spacecraft by the Sun and the cost of tracking the Sun's reflex motion. 

Because artifact SETI normally goes to great lengths to avoid assuming that an ETI is coincidentally investigating the Solar System just as humanity begins to explore space, it seems reasonable to conclude that an artifact SETI search for a stellar relay around the Sun depends on propulsion technologies more advanced than those available to humanity, or else send a continuous stream of replacement spacecraft as each exhausts it propellant. The observational characteristics of extremely hypothetical propulsion systems are of course equally unknown; for example, perhaps it is possible to detect a continuously firing antimatter reactor in nearby interstellar space.

\section{Alternative Stationkeeping Schemes}

\subsection{Solar Sails}

Besides traditional rocket technology, several technology demonstrators and experiments have begun using solar sail technology to use momentum from the Sun's photon emission to provide low but constant accelerations on deep-space missions. The acceleration $\mathcal{A}$ of a solar sail of area $A_s$, mass $m_s$ and reflective coefficient $k_s$ depends on the luminosity of the Sun $L_\odot$ and the distance $d$ from the Sun, plus the angle the sail forms with the Sun's rays $\theta$.

\begin{equation} \label{eq:sail}
    \mathcal{A} = \frac{L_\odot}{4 \pi d^2 c} \frac{ (1+k_s) A_s}{m_s} \cos^2{\theta}
\end{equation}

Because the gravitational force of the Sun follows a similar inverse-square law with distance, it is simple to determine the physical qualities of the sail materials needed to resist the inwards force of gravity due to the Sun. Setting the acceleration of the sail equal to the acceleration due to the Sun's gravity gives an expression for the required mass-to-area ratio $\frac{m_s}{A_s}$ such that the spacecraft does not need to spend propellant to resist the Sun's gravity. Assuming that the sail has perfect reflectivity $1+k_s = 2$ and that $\cos^2{\theta} = 1$ for directly outward thrust, equation \ref{eq:sail} becomes

\begin{equation*}
    \frac{G M_\odot}{d^2} = \frac{L_\odot}{2 \pi d^2 c} \frac{A_s}{m_s}
\end{equation*}

\begin{equation}
    \frac{L_\odot}{2 \pi c G M_\odot} = \frac{m_s}{A_s}
\end{equation}

Calculating this limiting sail density for resisting the Sun's gravity gives $m_s/A_s < 1.5$ $\rm{g/m^2}$. By comparison, gold leaf of thickness $0.1 \:\rm{\mu m}$ has a density of $1.9 \:\rm{g/m^2}$, suggesting that a relay spacecraft equipped with a sail with density comparable to gold leaf would still need traditional thrusters to maintain its position. The reflex motion of the planets is even more difficult to match with a solar sail, as the sail would need to turn away from normal to the Sun to gain perpendicular acceleration, dramatically decreasing $\cos^2{\theta}$. Furthermore, the acceleration due to the reflex motion is over ten times more than the Sun's inwards gravity. By these measures, if a solar sail is to contribute to the stationkeeping of a relay spacecraft, it must be very large, very reflective, and very lightweight (but not impossibly so) so solar sails may be a valuable component in countering some of the perturbations on the relay-Sun-target alignment, limited only by the endurance of the physical components of the sail in the face of impacts from interplanetary and interstellar material \citep{Bialy2018}. Such a sail may be visible from Earth if it is expansive and reflective enough, though its observability is difficult to predict \textit{a priori}. Alternatively, propellant-less navigation can be achieved via electric- or magnetic-field sails as discussed in \cite{Lingam2020}, which have several advantages over photon sails in deep space.

\subsection{Nonstationary probes}

Because the SGL focus is a line, not a point, a probe does not need to remain stationary, but could instead coast along the focal line opposite the target star using no radial propulsion at all.  If the spacecraft is gravitationally bound to the Sun, this could result in long periods of transmission followed by periodic trips into the inner Solar System. Such a probe  would still need to expend propulsion to move laterally to avoid a close approach with the sun once per orbit, and then again to come back into alignment on the outbound trip.

This scheme would still need to pay essentially all of its stationkeeping costs in the high-gain case where it adjusts for the Sun's motion, but in the low-gain case the fuel cost is somewhat lower, because the fuel needed to perform this maneuver is less than is needed to resist the Sun's gravity alone. To ``miss'' the Sun on the way in, the probe would need to give itself sufficient angular momentum to achieve periherlion at some distance $r_p$.  Since it is barely bound, we can estimate this specific angular momentum as $l = r_p v_\mathrm{esc}(r_p)$. If we assume it remains on the focal line until the minimum SGL focus distance of 550 AU, then it requires only
\[
\Delta v = \frac{r_p}{550\ \mathrm{au}}\sqrt{\frac{2GM_\odot}{r_p}} = 80 \mathrm{\ m\ s}^{-1} \sqrt{\frac{r_p}{\mathrm{au}}}
\]

\noindent to acquire sufficient lateral motion, which it must spend again to correct on its way out. If we assume an orbital period of 11,000y (giving it an aphelion distance of 1000 au) then the time-averaged acceleration for perihelion distance of 1 au is 0.016 m/s/y, which is smaller than the acceleration due to the Sun at 550 au by a factor of 3.  This number can be improved further with closer approaches to the sun or larger aphelion distances.  So, such a scheme is somewhat better than simple stationkeeping, but comes at the cost of a loss of duty cycle.

The lower duty cycle could be overcome by having multiple probes, and indeed occasionally visiting the inner solar system might be a goal of such probes. This might provide motivation to explore high-eccentricity objects with periastron values beyond 550 au near the antipodes of nearby stars for evidence that such objects are artificial.

\subsection{Stationary fuel, nonstationary probes}

Our use of Eq.~\ref{RocketEquation} assumes that the probe carries its fuel with it. An alternative scheme would be to have the fuel remain in a stationary fuel depot, fighting only the inward pull of gravity, so that a high gain probe could maintain alignment more efficiently by only carrying a small amount of fuel at any time. In such a scheme additional fuel could be transported to the probe occasionally at negligible fuel cost when the probe and depot have low mutual velocity. The total lifetime of the probe is then simply and approximately

\[
\tau \sim \frac{v_\mathrm{ex}}{A_\mathrm{ref}}\frac{m_\mathrm{fuel}}{m_e}
\]

\noindent where, again, $m_e$ is the dry mass of the probe. This compares favorably by a factor of a few to Eq.~\ref{Lifetime} when $m_\mathrm{fuel} = m_e$ and improves quickly as the ratio $m_\mathrm{fuel}/ m_e$ grows, suggesting this is a superior scheme than the one we describe in earlier sections. The benefit is limited, however: in the limit of large amounts of fuel, the lifetime is set by the stationkeeping of the fuel stores, not the probe, and so such schemes would have maximum lifetimes of order a few times $\tau_\mathrm{max}$ in Table~\ref{tab:Spaceships}, even for high gain communication.

\section{Summary}
\label{sec:Summary}

In examining the engineering of a long-lived stellar gravitational lens relay, the substantial costs imposed by dynamical perturbations on the relay-Sun-target alignment come into view.  These considerations have implications for the construction of stellar relays (whether solitary or as part of a galactic network) and for artifact SETI searches for relay spacecraft using the Sun's gravitational lens.

From an engineering standpoint for a stellar relay around the Sun, our conclusions are...

\begin{itemize}
    \item Though the on-axis gain of a stellar relay system around the Sun can reach $>100 \:\rm{dB}$ when transmitting at sub-mm wavelengths, these high gains require a near-perfect trifold relay-Sun-target alignment maintained to $< \rm{km}$ precision.
    \item Ignoring all other concerns, the relay spacecraft must expend delta-v around $0.5\:\rm{m/s/yr}$ to resist the Sun's inward gravity. This places an upper limit on the lifetime of a stellar relay with finite available delta-v.
    \item Planetary companions impart significant reflex motions on a host star, which must be actively corrected by the relay spacecraft to maintain high system gains.  For example, Jupiter impacts a reflex acceleration of $6.6 \:\rm{m/s/yr}$ on the Sun, which increases the delta-v stationkeeping cost a relay spacecraft must pay to maintain alignment for an interstellar communication.
    \item The reflex motions on stars in close binary systems like Sirius or $\alpha$ Cen are dramatically higher than in single star systems, rendering such systems much less economical for stellar relays of any type or purpose. Multiple-star systems are only economical for stellar relays if the orbital distance between the two stars is much greater than the minimum focal distance for the lensing star.
    \item Alternative methods of maintaining focus at the SGL might improve the maximum lifetimes we calculate for the Sun by a factor of a few for low gain probes (by moving along the focus instead of strict stationkeeping) and up to the low-gain probe limit for high-gain probes (by storing fuel separately from the probe itself).
\end{itemize}

From an artifact SETI perspective, our conclusions are...

\begin{itemize}
    \item Artifact SETI searches for ETI-placed relay spacecraft around the Sun are only sensitive to interstellar transmissions from that spacecraft if the opening angle of the initial unfocused beam is large enough to include Earth's orbit. This places an upper limit on the spacecraft transmission gain and dish size; otherwise, an artifact SETI search depends on the relay spacecraft transmitting in other ways.
    \item Some star systems are more or less economical for hosting a stellar relay than others. A node-based galactic communications system might exclude many star systems based on the perturbations of the gravitational lenses therein, or the degree to which the gravitational lens adheres to a spherical mass distribution.
    \item The Sun, while lacking close stellar companions, is moderately perturbed by its gas giant planets. There may be other star systems nearby more hospitable to long-lived stellar relays.
    \item To sustain a stellar relay around the Sun for millions of years, an ETI must use advanced propulsion technology far more energetic than human systems like chemical, nuclear thermal, or even ion rockets. Alternatively the ETI may simply replace the relay spacecraft every thousand years or so.
\end{itemize}

Based on our analysis, we conclude that it is feasible to conduct observational searches for ETI-placed spacecraft transmitting using the SGL. These searches are predicated on the spacecraft still actively transmitting and maintaining its position in the face of the perturbations noted in Section \ref{sec:Perturbations}. Future searches might focus on antipodes of nearby star systems that meet our criteria for dynamic stability and usefulness for stellar gravitational lens transmission.  As the transmission from an ETI-placed relay are not necessarily intended for interception, we cannot make \textit{a priori} guesses about the optimal frequencies of such SETI searches, but we note that if a spacecraft is actively maintaining its position it may be observable due to by-products of this propulsion in addition to its interstellar communications.

\section*{Acknowledgements}

This research started as a final project for the 2020 graduate course Astro 576: ``The Search for Extraterrestrial Intelligence" at the Pennsylvania State University Deptartment of Astronomy and Astrophysics. This research developed with valuable discussion helpfully given by Cayla Dedrick, Laura Duffy, Gregory Foote, Macy Huston, Andrew Hyde, Julia Lafond, Ella Mullikin, Michael Palumbo, Winter Parts, Phoebe Sandhaus, Hillary Smith, Evan Sneed, and Nick Tusay.

Endless thanks is due for comments on drafts of this work by Michael Hippke and Alexander Madurowicz.  We also thank the anonymous reviewer of this work for their comments, guidance, and suggestion.

The Center for Exoplanets and Habitable Worlds and the Penn State Extraterrestrial Intelligence Center are supported by the Pennsylvania State University and its Eberly College of Science.

\bibliography{main}{}

\end{document}